\documentclass{ws-rv975x65}
\usepackage{subfigure}     
\usepackage{ws-rv-van}     
\makeindex
\begin{document}

\chapter[Circularly Polarized Attosecond Pulses and Molecular Atto-Magnetism]{Circularly Polarized Attosecond Pulses and Molecular Atto-Magnetism}\label{ra_ch1}

\author[F. Author and S. Author]{Andr\'{e} D. Bandrauk \footnote{Canada Research Chair - Computational Chemistry and Molecular Photonics} and Kai-Jun Yuan}

\address{Laboratoire de Chimie Th\'{e}orique, Facult\'{e} des Sciences, Universit\'{e} de Sherbrooke, Sherbrooke, Qu\'{e}bec, Canada, J1K 2R1, \\
andre.bandrauk@usherbrooke.ca}

\begin{abstract}
Various schemes are presented for the generation of circularly polarized molecular high-order harmonic generation (MHOHG) from molecules. In particular it is shown that combinations of counter-rotating circularly polarized pulses produce the lowest frequency Coriolis forces with the highest frequency recollisions, thus generating new harmonics which are the source of circular polarized attosecond pulses (CPAPs). These can be used to generate circularly polarized electronic currents in molecular media on attosecond time scale. Molecular attosecond currents allow then for the generation of ultrashort magnetic field pulses on the attosecond time scale, new tools for molecular atto-magnetism (MOLAM).
\end{abstract}

\body

\section{Introduction}\label{ra_sec1}

Attosecond science opens the door to real time observations and control of electron dynamics on the electron's natural time scale, the attosecond (1 asec= 10$^{-18}$ s) \cite{1,2,3}. The study of ultrafast electron motion in matter from molecules to materials is a new frontier of modern science due to the rapid evolution of laser technology, allowing for the synthesis and even shaping of ever shorter (few cycles) and more intense laser pulses. Since 152 asec is the classical period of revolution of the 1s electron in the ground state of the H atom, increasing laser intensity and/or nuclear charge as in the U atom ($Z=92$), the energy band width $\Delta E =mc^2$ necessary to cover the electron (e$^-$)-positron (e$^+$) threshold for the creation of antimatter requires pulse durations of $\hbar/mc^2\simeq1$ zeptosecond (1 zps=10$^{-21}$ s) \cite{4}. It is through the interaction of current ultrashort intense pulses of intensity $I_0\geq3.5\times10^{16}$ W/cm$^2$, corresponding to an electric field strength $E_0\geq5.14\times10^9$ V/cm, the atomic units (a.u.) of field intensity $I_0$ and field strength $E_0$, that one enters a new regime of laser-matter interaction, the highly nonlinear, nonperturbative regime where the electric $\textbf{E}$-magnetic $\textbf{B}$ fields of intense pulses control electron-nuclear motion \cite{5}. This regime allows for the generation of new coherent attosecond pulses for monitoring and controlling electrons in molecules \cite{3,6} and zeptosecond pulses for nuclear fusion \cite{7,8}.

The field of attosecond science is a new emerging science requiring improvement and development via theory and high level simulations the quality and stability of attosecond and even zeptosecond pulses. Quantum numerical simulations have been at the forefront of this research developing these new tools for tackling also photochemistry, photobiological and molecular photonics problems \cite{9}. Molecular high-order harmonic generation (MHOHG) is now used to monitor electron dynamics in molecules \cite{3,5,10}, to probe electrons at surface \cite{11}, and to transport electron coherently to large distance \cite{12} on attosecond-femtosecond time scales. The synthesis of attosecond pulses relies on the time-energy uncertainty principle, $\Delta E \Delta t \geq \hbar$, i.e., collecting together light coherent sources with an energy bandwidth $\Delta E$ gives a pulse with duration $\tau \simeq \hbar/\Delta E$. The shortest pulse, a delta function time pulse $\delta(t)$ used to imaging molecular orbitals \cite{13} illustrates this simple principle:
\begin{equation} \label{eq1}
\delta(t)=\frac{e^{i\phi}}{2\pi} \int_{-\infty}^{\infty}e^{i\omega t} dt.
\end{equation}
Of note is that the strength of all electric field amplitudes $e^{i\omega t}$ is equal in addition to the phase $e^{i\phi}$. To date the most convenient source of such electric field amplitudes is high-order harmonic generation (HHG) for atoms and MHOHG in molecules. Furthermore, to date HHG and/or MHOHG are obtained solely using linearly polarized excitation electric pulses $E(t)=E_0f(t)\cos(\omega t +\phi)$, where $f(t)$ is the pulse envelope, $\phi$ the carrier envelope phase (CEP), and $E_0$ the electric field amplitude which allows to define the maximum intensity $I_0=cE_0^2/8\pi$.

Such linearly polarized pulses imply recollision of ionized electron with the parent ion \cite{14,15}, thus allowing to predict the maximum energy and order $N$ in HHG as: \begin{equation} N\hbar \omega=I_p+ 3.17U_p, \end{equation} where $U_p=I_0/4m_e\omega^2$, is the pondremotive energy and $I_p$ is the ionization potential, previously obtained in numerical simulations \cite{16}. As pointed out by Corkum, for circularly polarized light, electron trajectories never return to the vicinity of the ion and electron-ion interactions are not important for such polarization \cite{14}. Nevertheless circular polarization ionization in the long wavelength limit results in broad electron classical energy distributions, called above threshold ionization (ATI) spectra, with peaks at $2U_p$ \cite{17}. The general conclusion from such early work was that whereas the width of the energy distribution for linearly polarized pulses is determined by the phase angle $\phi$ at which the electron is ionized, the energy width for circularly polarized light is characteristic of the pulse envelope $f(t)$ \cite{14}.

Recent classical nonlinear dynamical theory and simulations show that in strong circularly polarized laser fields, key periodic classical orbits can drive recollisions due to Coulomb potentials \cite{18}. Nonadiabatic theory of strong field atomic ionization has also shown that recollision and correlated double and triple ionization are found to be possible with elliptical polarization \cite{19}. Atomic ionization by strong elliptically polarized laser pulses has also been studied analytically and numerically \cite{20,21,220} based on the original semi-classical model of Keldysh \cite{22}. Contrary to atoms, molecules offer the opportunity of examining the effect of Coulomb multi-center effects on strong field ionization and laser induced recollision. One important manifestation of such effects is laser induced electron diffraction (LIED) \cite{13} by long pulses and recently by linearly and circularly polarized attosecond pulses \cite{23}. MHOHG differs dramatically from atomic HHG by the emission of elliptically polarized MHOHG spectra even with linearly polarized laser pulses due to the nonspherical, but cylindrical symmetry in diatomic molecules \cite{24}. Furthermore attosecond circular dichroism is inherent in MHOHG of aligned molecules due to multiple molecular electronic continua \cite{25}. The effect of nuclear motion on MHOHG and on generation of attosecond pulses in linearly polarized intense laser pulses obtained from non-Born-Oppenheimer numerical solutions of the time-dependent Schr\"{o}dinger equation (TDSE) for one dimension (1D) H$_2$ has shown that nuclear motion shortens attosecond pulse trains originating from the first electron ionization \cite{26}. Another important difference between molecular and atomic HHG is the possibility in linear polarization for the ionized electron to recombine with neighboring ions at large internuclear distance thus extending MHOHG plateaus beyond the atomic 3.17$U_p$ cut-off law [Eq. (2)] to 8$U_p$ \cite{15,27,28}.

Strong field electron emission from aligned H$_2^+$ ions with circularly polarized laser pulses has shown complex laser driven electron dynamics due to the two-center Coulomb potentials \cite{29}. Another source of the circularly polarized HHG spectra is ring current initial states \cite{30} and non-zero initial velocity electrons \cite{31}. In general, since circularly polarized laser pulses create electrons with ``spinning" electron trajectories with very large radii, recollision with parent ions is very unlikely in atoms and molecules at equilibrium, except at large internuclear distance \cite{32}, thus making circularly polarized HHG very inefficient. Alternative, solutions to induce recollision with circularly polarized pulses have been proposed using bicircular fields \cite{33,33a,34}. Milo\v{s}evi\'{c} and Becker have analyzed the case of superposition of two coplanar counter-rotating circularly polarized fields. They have shown that superposing a few cycle circularly polarized pulse and a long circularly polarized counter-rotating pulse can generate a field that is linearly polarized for extremely short time and this can be used to produce single linearly polarized attosecond pulses \cite{34} in atomic media. We have also studied MHOHG, i.e., in molecular media, and have shown that single circularly polarized pulses with time modulations of the envelope $f(t)$ gives rise to two different frequency circularly polarized pulses and always lead to recollision with the molecular plane \cite{35,36}. This is achieved by studying the laser induced electron dynamics in a rotating frame, which separates the dynamics under the influence of a \textit{Coriolis} force and the  \textit{recollision} potential. In both papers it is shown that the Coriolis force can be controlled by a static magnetic field of strength $B$ (gauss). The effect of static magnetic and electric fields has also been examined theoretically for linear laser polarization with the condition that such a scenario will extend the maximum HHG energy cut-off beyond $3.17U_p$ \cite{37}. Using the ideas proposed previously by us \cite{35,36}, we have recently shown new methods for generating circularly polarized MHOHG \cite{32,38}, circularly polarized attosecond pulses \cite{39}, and attosecond magnetic field pulses \cite{40}. We note that other previous methods to produce sources of circularly polarized HHG relied on reflector phase-shifters of linearly polarized infrared laser light with an efficiency of a few percent, thus proposing such new pulses as new tools for the investigation of ultrafast magnetization dynamics \cite{x,y}. In the next section we present and expand the theoretical ideas which propose a new science based on circularly polarized attosecond laser pulses: Atto-Magnetism.

\begin{figure}[!t]\centering
\includegraphics[scale=1,angle=0]{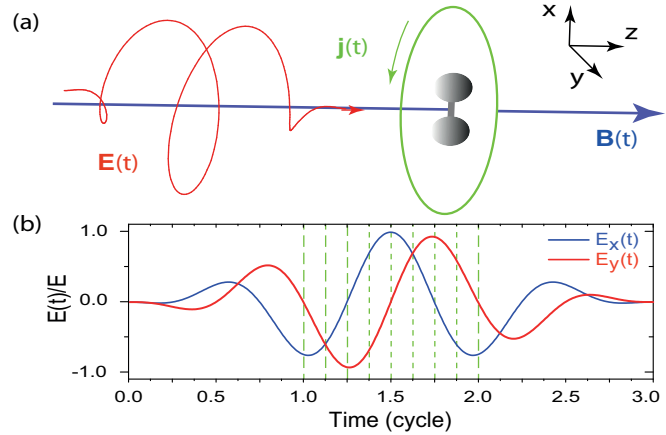}
\caption{(a) Illustration of the attosecond magnetic fields $\textbf{B}(\textbf{r},t)$ (blue line, along the $z$ axis) for H$^+_2$ created by a few cycle circularly polarized attosecond UV pulse (red line). The green line represents the corresponding current $\textbf{j}(\textbf{r},t)$ in the molecular ($x,y$) plane. The magnetic field $\textbf{B}(\textbf{r},t)$ is perpendicular to the current $\textbf{j}(\textbf{r},t)$. (b) Three cycle circularly polarized attosecond UV pulse $\textbf{E}(t)$ at $\lambda=50$ nm. Dashed lines correspond to the times in Table \ref{tab1}.  }
\label{fig2}
\end{figure}

\section{Theoretical Molecular Model}

We begin with the exact three-dimension (3D) Hamiltonian of H$_2^+$ with static (Born-Oppenheimer) nuclei adapted to the magnetic field $B$ (gauss)
problem described previously \cite{35,41} (we use atomic units: $e=\hbar=m_e=1$),
\begin{equation} \label{eq2}
i\frac{\partial \psi(x,y,z,t)}{\partial t}= \displaystyle \left [ H_0 +H_l+H_B \right ] \psi(x,y,z,t),
\end{equation}
where
\begin{eqnarray} \label{eq3}
H_0 &=& \displaystyle \frac{2m_p+1}{4m_p}\left [ \frac{\partial^2 }{\partial x^2} +\frac{\partial^2 }{\partial y^2} + \frac{\partial^2 }{\partial z^2} \right ]- \frac{1}{\sqrt{(x\pm R/2)^2+ y^2+z^2}},
\end{eqnarray}

\begin{eqnarray} \label{eq4}
H_l &=& \kappa E_0 \cos (\omega t) [x\cos(\bar \omega t) + y \sin (\bar \omega t)] \nonumber \\
&=&  \kappa \frac{E_0}{2}\{x\cos[(\omega + \bar \omega) t]+y \sin[(\omega + \bar \omega) t]\} \nonumber \\
&& + \kappa \frac{E_0}{2}\{x\cos[(\omega - \bar \omega) t]-y \sin[(\omega - \bar \omega) t]\},
\end{eqnarray}

\begin{eqnarray} \label{eq5}
\begin{array}{c }
H_B = \displaystyle  \beta l_z +\frac{1}{2} \beta^2\rho^2-\beta, \\ \displaystyle l_z=-i\left [ x\frac{\partial}{\partial y} - y\frac{\partial}{\partial x} \displaystyle \right ] ,\\ \rho=\sqrt{x^2+y^2}.
\end{array}\end{eqnarray}
$\kappa=(2m_p+2)/(2m_p+1)$, $m_p=1837$ is the mass of the proton, $\beta=B/B_0$, and $B_0=2.35\times10^9$ G, the atomic unit of the magnetic field. We neglect here the effect of the magnetic field on the nuclear motion \cite{42} since the resulting Lorentz force scales as $v/c=1/137$ for one atomic unit (a.u.) of velocity and should therefore be negligible for the proton whose mass is 1837 a.u. Our model, Fig. \ref{fig2}(a) therefore consists of H$_2^+$ aligned with the $x$ axis in the presence of a static magnetic field of strength $B$ parallel to the internuclear axis $R$. The magnetic field induces angular momentum $l_z$ perpendicular to the $x$ molecular axis and a confining potential $\beta^2\rho^2/2$ around that axis (the diamagnetic energy). The parameter that characterizes the magnetic confinement is the Landau radius due to harmonic motion in that confining potential, i.e., $R_L=\sqrt{2/\beta}$. Thus a magnetic field of strength $\beta=0.1$ (2.35$\times10^8$ G) will induce a Landau radius $R_L\simeq 4.5$ a.u. in the $(x,y)$ molecular plane (Fig. \ref{fig2}). The modulated field in Eq. (\ref{eq4}) corresponds to two circularly polarized fields which differ by $2\bar \omega$ in frequency, co-rotating at frequency $\omega+\bar \omega$ and counter-rotating at frequency $\omega-\bar \omega$.

Applying the unitary transformation $\mathcal T=\exp(-i\bar \omega t l_z)$, which is a rotation around the $+z$ direction by the angle $\bar \omega t$,
gives the new Hamiltonian
\begin{equation} \label{eq6}
H'=H_0+\frac{1}{2}\beta^2\rho^2+(\beta-\bar \omega )l_z +xE_0\cos(\omega t).
 \end{equation}
The new Hamiltonian exhibits a magnetic confinement potential $\frac{1}{2}\beta^2\rho^2$, a rotation term $(\beta-\bar \omega )l_z$, and a linear driving field term $xE_0\cos(\omega t)$ in the frame rotating at frequency $\bar \omega $. This last term will give rise to a ponderomotive energy $U_p$. The model of MHOHG via recollision of the ionized electron with the ion core predicts a maximum harmonic order $N$, Eq. (2). We note here that since $x$ is not diagonal in the angular momentum quantum number $m$ (around the $z$ axis), the last laser-driving term in Eq. (\ref{eq6}) couples different angular momentum terms, thus also pumping energy into circular motion.

Equations (\ref{eq2}-\ref{eq6}) demonstrate a universal behavior of modulated envelope circularly polarized pulses. Considering zero magnetic fields ($\beta=0$), then at $\bar\omega= \omega$, Eq. (\ref{eq4}) becomes a single circularly polarized field of frequency $2\omega$ in the presence of a static electric field of strength $E_0/2$, which simplifies Eq. (\ref{eq6}) to
\begin{equation} \label{eq7}
H'=H_0-\omega l_z+xE_0\cos(\omega t),
\end{equation}
i.e., a Coriolis force of the same frequency $\omega$ as the recollision electric field but half that of the incident circular pulse. We have used such a pulse combination to obtain circularly polarized MHOHG in H$_2^+$ \cite{38}. Setting $\omega\sim \bar \omega$ gives a combination of a fast $\omega+\bar \omega$ circularly polarized pulse with a nearly static TeraHertz field of frequency $(\omega-\bar \omega)$. Equation (\ref{eq6}) then reduces at $\beta=0$ to
\begin{equation} \label{eq8}
H'=H_0-\bar \omega l_z+xE_0\cos(\omega t).
\end{equation}
This guarantees weaker Coriolis forces of lower frequency $\bar \omega$ in the presence of a strong linear ionizing field of higher frequency $\omega$. The effect of the strong field $E_0$ is to force recollision and produce a circularly polarized MHOHG spectrum \cite{39}. It is to be noted that a single circularly polarized pulse results in the rotating frame with the Hamiltonian at $\omega=0$, i.e., a Coriolis force with a static electric field $E_0$ only, Eq. (\ref{eq7}). Such a Hamiltonian has been previously studied for Rydberg states in microwave fields, creating static classical periodic orbits \cite{43}.

Static electric field induced polarizations have been studied in harmonic generation with the general condition that such fields result in elliptic dichroism in which the harmonic yield is different for right and left elliptically polarized laser fields \cite{44}. In circularly polarized atomic HHG and molecular MHOHG, one requires the harmonic field components $E_x=E_y$ and maintain a phase difference of $\pi/2$ \cite{38,39,40} due to recollision, involving therefore complex laser electron recollision dynamics \cite{32}. Equation (\ref{eq4}) clearly predicts two different bicircular combinations. For $\omega>\bar \omega$, one obtains two \textit{counter-rotating} circularly polarized laser pulses whereas for $\omega<\bar \omega$, one obtains two \textit{co-rotating} pulses. Equation (\ref{eq6}) predicts for all two cases a strong linearly polarized pulse which will induce recollision at frequency $\omega$ and a corresponding ponderomotive energy $U_p$. We define next different frequency ratios $\bar \omega/\omega$ which will lead to various possible schemes for creating HHG or MHOHG, using Eqs. (\ref{eq4}) and (\ref{eq6}). Thus setting $\bar \omega=2\omega$, one obtains neglecting magnetic fields ($\beta=0$),
\begin{eqnarray} \label{eq9}
E(t)&=&\displaystyle \frac{E_0}{2}\left[ x\cos(3\omega t)+y\sin(3 \omega t) \right] \nonumber \\
&& \displaystyle + \frac{E_0}{2}\left[ x\cos(\omega t)+y\sin( \omega t) \right] ,
\end{eqnarray}

\begin{eqnarray} \label{eq10}
H'(t)=H_0-2\omega l_z +xE_0\cos(\omega t).
\end{eqnarray}
Such a $3\omega$ and $\omega$ combination of co-rotating circularly polarized pulses results in recollision at frequency $\omega$ but with a Coriolis force with $2\omega$ frequency.

Setting $\omega=2 \bar \omega$ leads to $3 \bar \omega$ and $ \bar \omega$ counter-rotating circularly polarized pulses.
\begin{eqnarray} \label{eq11}
E(t)&=&\displaystyle \frac{E_0}{2}\left[ x\cos(3\bar \omega t)+y\sin(3 \bar\omega t) \right] \nonumber \\
&& \displaystyle + \frac{E_0}{2}\left[ x\cos(\bar \omega t)-y\sin(\bar \omega t) \right],
\end{eqnarray}

\begin{eqnarray} \label{eq12}
H'(t)=H_0-\bar \omega l_z +xE_0\cos(2 \bar\omega t),
\end{eqnarray}
with now the recollision frequency $2 \bar \omega$ is twice that of the Coriolis force. Comparing Eqs. (\ref{eq10}) and (\ref{eq12}), we note the large Coriolis frequency in the co-rotating system, Eq. (\ref{eq9}).

Setting $\bar \omega=3 \omega$ leads to $4  \omega$ and $ 2 \omega$ co-rotating circularly polarized pulse scheme,
\begin{eqnarray} \label{eq13}
E(t)&=&\displaystyle \frac{E_0}{2}\left[ x\cos(4 \omega t)+y\sin(4\omega t) \right] \nonumber \\
&& \displaystyle + \frac{E_0}{2}\left[ x\cos(2 \omega t)+y\sin(2 \omega t) \right],
\end{eqnarray}

\begin{eqnarray} \label{eq14}
H'(t)=H_0-3 \omega l_z +xE_0\cos(\omega t),
\end{eqnarray}
where the Coriolis force acts at three times the frequency of the recollision force.

Setting $\omega=3 \bar \omega$ leads to a $4 \bar \omega$ and $ 2 \bar \omega$ counter-rotating scheme with the recollision force at frequency $3 \bar \omega$ whereas the Coriolis force has frequency $\bar \omega$ only. This scheme has been investigated in detail \cite{33,33a,34} but not in the rotating frame. Finally setting $2(\omega+\bar \omega)=3(\omega-\bar \omega)$, which corresponds to $\omega=5 \bar \omega$ leads to counter-rotating pulses with relative frequencies $3\omega$ and $2 \omega$, i.e., with $\omega=2 \bar \omega$,
\begin{eqnarray} \label{eq15}
E(t)&=&\displaystyle \frac{E_0}{2}\left[ x\cos(6\bar \omega t)+y\sin(6 \bar\omega t) \right] \nonumber \\
&& \displaystyle + \frac{E_0}{2}\left[ x\cos(4 \bar \omega t)-y\sin(4 \bar \omega t) \right],
\end{eqnarray}

\begin{eqnarray} \label{eq16}
H'(t)=H_0-\bar \omega l_z +xE_0\cos(5 \bar\omega t),
\end{eqnarray}
whereas the co-rotating combination for $\bar \omega=5 \omega$
\begin{eqnarray} \label{eq17}
E(t)&=&\displaystyle \frac{E_0}{2}\left[ x\cos(6  \omega t)+y\sin(6  \omega t) \right] \nonumber \\
&& \displaystyle + \frac{E_0}{2}\left[ x\cos(4  \omega t)+y\sin(4   \omega t) \right],
\end{eqnarray}

\begin{eqnarray} \label{eq18}
H'(t)=H_0-5 \omega l_z +xE_0\cos( \omega t).
\end{eqnarray}

In general setting $\omega=n\bar \omega$ leads to counter-rotating pulses with relative frequencies $(n+1)\bar \omega$ and $(n-1)\bar \omega$, Eq. (\ref{eq4}) which in the rotating frame with frequency $\bar\omega$, gives Coriolis forces $\bar \omega l_z$ and recollision electron field $E_0 \cos (n\bar \omega t)$, Eqs. (\ref{eq12}) and (\ref{eq16}).

\begin{figure}[!t]\centering
\includegraphics[scale=0.95,angle=0]{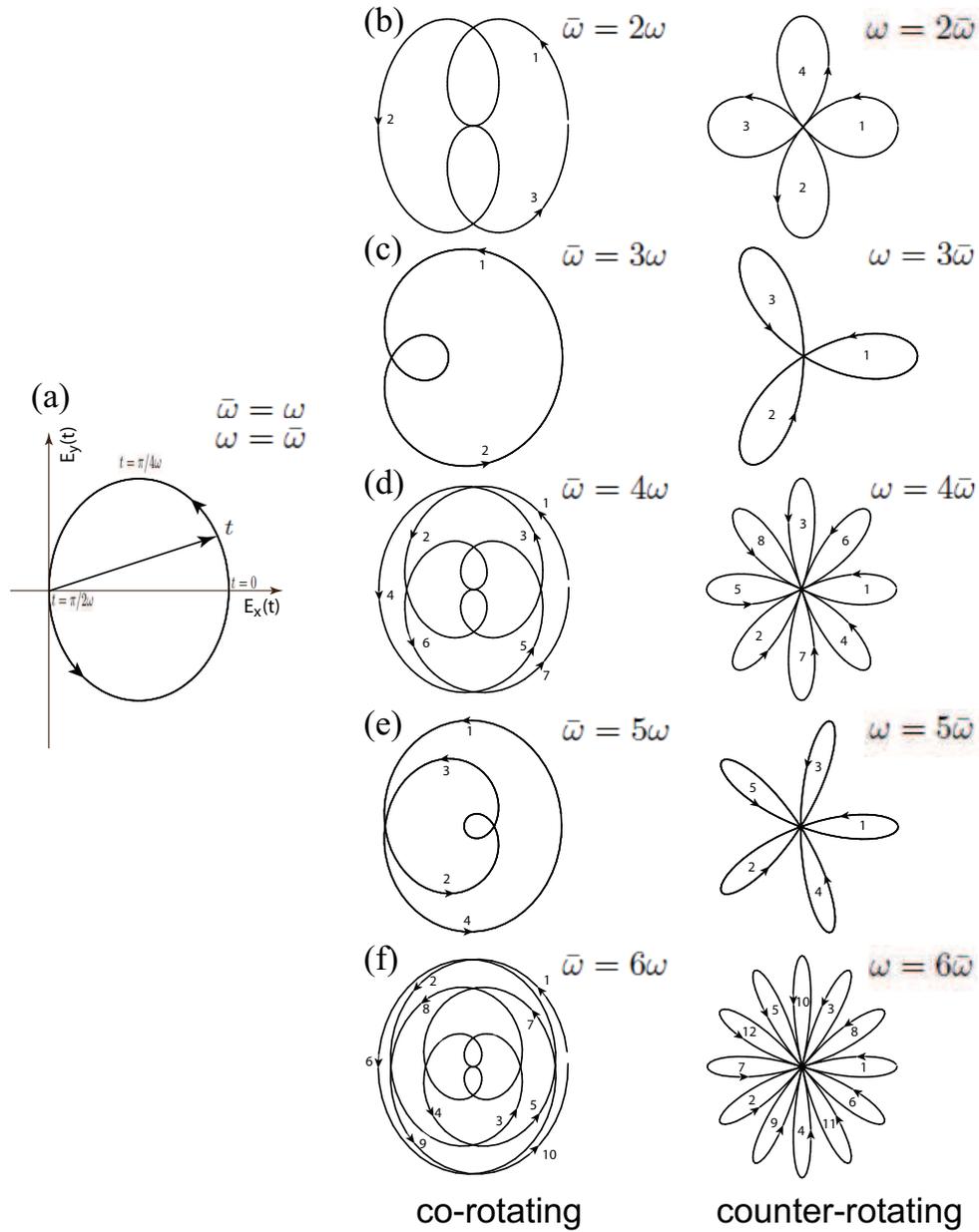}
\caption{The electric field vector $\textbf{E}(t)$ Eq. (\ref{eq4}) for the various ($\omega,\bar \omega$) combination schemes. Panel (a): For $\omega=\bar \omega$, the electric field is a superposition of a circularly polarized pulse and a static field \cite{35,36}. Panels (b-f): For co-rotating combinations $\bar \omega = n \omega$ and counter-rotating combinations $\omega = n \bar \omega$. Arrows indicate the evolution of the fields with time $t$.}
\label{fig1}
\end{figure}
 Equations (\ref{eq9}-\ref{eq18}) confirm that for equal intensity $E_0$ in co-rotating and counter-rotating pulse combinations, Eq. (\ref{eq4}), it is the co-rotating schemes, Eqs. (\ref{eq9}), (\ref{eq13}), and (\ref{eq17}) which always produce large frequency Coriolis forces $\omega l_z$ accompanied by lower frequency recollision forces whereas the counter-rotating schemes,  Eqs. (\ref{eq11}) and (\ref{eq15}) lead to lower frequency Coriolis forces accompanied by much higher frequency recollisions. We illustrate next the net electric vectors in Fig. \ref{fig1} for the various schemes described above in addition to the counter-rotating combinations $\omega=4\bar \omega$ and $6 \bar \omega$ and the co-rotating combinations $\bar \omega=4 \omega$ and $6  \omega$. We note that for all the counter-rotating combinations $\omega=n\bar \omega$, the net electric field includes a zero field amplitude thus maintaining an inversion symmetry. The co-rotating field combinations $\bar \omega = n \omega$ show only a reflection symmetry with respect to the $x$ axis. A similar behavior has been found for a ($\omega,2\omega$) combination of equal amplitude circular polarizations \cite{34,z}. This corresponds to the $\bar \omega =3\omega$ and $\omega=3 \bar \omega$ combinations in Fig. \ref{fig1}. These authors have also concluded that only one quantum orbit contributes to the harmonics in this case. However the HHG spectrum did not show a continuous circularly polarized spectrum.

 The case $\omega=\bar \omega$ as mentioned above leads to the superposition of a circularly polarized pulse and a static field which in the rotating frame of frequency $\omega$, half the frequency of the incident circularly polarized pulses, leads to a Coriolis force of the same frequency $\omega$ as the recollision force, Eq. (\ref{eq7}). It was found that such a combination leads to a circularly polarized MHOHG plateau from which one obtained $\sim 100$ asec circularly polarized pulses \cite{38}. The intensities required for attosecond pulse generation was $I\geq10^{14}$ W/cm$^2$ for both circularly polarized pulse and static fields. Since TeraHertz pulses can approach such intensities and thus generate nearly static fields, numerical simulations of the TDSE for H$_2^+$ showed that attosecond circularly polarized pulses were also generated under similar conditions as the static field scheme \cite{39}. Time-series analysis of the circularly polarized MHOHG spectra confirmed the generation of the harmonics by single recollisions with the parent ion as opposed to the linear polarization excitation spectra \cite{15}.

 Equation (\ref{eq6}) shows that Coriolis effects can be controlled by the introduction of an external magnetic field $\beta=B/B_0$ (a.u.). The magnetic field furthermore introduces a magnetic confinement potential $\frac{1}{2}\beta^2\rho^2$ in the molecular ($x,y$) plane (Fig. \ref{fig1}), electron density. Solving exactly the 3D TDSE for the H$_2^+$ molecular ion allowed for the study of the MHOHG spectrum control of electron recollision with the protons. Results were found of pure even or even and odd harmonic generation for particular field and frequency configurations. Furthermore magnetic fields were found to extend MHOHG plateaus to higher order \cite{36}.

\section{Attosecond Magnetic Field Pulse Generation}

Intense attosecond magnetic field pulses have been predicted to be produced by intense circularly polarized few cycle or attosecond ultraviolet(UV) laser pulses. Numerical solutions of the TDSE, Eqs. (\ref{eq2}-\ref{eq5}), yield exact time-dependent functions $\psi(\textbf{r},t)$ from which electron accelerations are calculated via the time-dependent Hellmann-Feynman theorem \cite{15},
\begin{equation} \label{eq20}
 \ddot{r}(t)=\langle \psi({r},t)|-\partial H/\partial r | \psi (r,t) \rangle -E(t).
\end{equation}
The MHOHG power spectrum $P_r(\omega)$ is then obtained from the absolute square of the Fourier transform (FT) of Eq. (\ref{eq20}),
\begin{equation} \label{eq21}
P_r(\omega)=|a_r(\omega)|^2=|\int \exp(-i\omega t) \ddot{r} (t) dt|^2.
\end{equation}
We note that the MHOHG spectra can be calculated numerically from the FT of the dipole moment $\langle r(t)\rangle$, velocity $\langle \dot r(t)\rangle$, and acceleration $\langle \ddot{r} (t) \rangle$ and must give identical results due to gauge invariance \cite{zz}. However it is found that the acceleration form Eq. (\ref{eq20}) gives more accurate results due to its emphasis on the recollision process where the force $-\partial H/\partial r$ is maximum. The temporal profile of the field components $\varepsilon_{x,y}(t)$ for a generated attosecond pulse is obtained further by an inverse FT of the MHOHG amplitudes $a_{x/y}(\omega)$ by,
\begin{equation} \label{eq22}
 \mathcal I_{x/y}(t)=|\varepsilon_{x/y}(t)|=|\int_{\omega_c} \exp(i \omega t) a_{x/y}(\omega)  d \omega|.
\end{equation}
The relative attosecond pulse phase difference $\varphi$ for circularly polarized pulses is obtained as
\begin{equation} \label{eq23}
\varphi (t)=|\arg [\varepsilon_{x}(t)]-\arg [\varepsilon_{y}(t)]| =\pi/2.
\end{equation}

Such new attosecond pulses obtained from circularly polarized MHOHG spectra according to Eqs. (\ref{eq20}) and (\ref{eq21}), we define now as
\begin{equation} \label{eq24}
E(t)=Ef(t)[\hat e_x\cos(\omega t)+\hat e_y\sin(\omega t)],
\end{equation}
propagating in the $z$ direction with $\hat e_{x/y}$ polarization directions. A smooth $n_l$ cycle $\sin^2(\pi t/n_l\tau)$ pulse envelope $f(t)$ characterizes such an attosecond pulse for maximum electric field amplitude $E$ and intensity $I=c \varepsilon_0 E^2/2$ and optical cycle $\tau=2 \pi/\omega$. Such a pulse satisfies a total area $\int E(t)dt=0$ \cite{15}. The TDSE at zero static magnetic field ($\beta=0$), Eq. (\ref{eq2}) is solved with the pulse, Eq. (\ref{eq24}), providing again the time dependent electron wavefunction $\psi(\textbf{r},t)$ for the pulse orientation illustrated in Fig. \ref{fig2}. Such laser-molecule interaction leads to an electronic current $\textbf{j}(t)$ and corresponding generated magnetic field $\textbf{B}(r,t)$
\begin{equation} \label{eq25}
 \textbf{j}(\textbf{r},t)=\frac{i}{2}[\psi(\textbf{r},t) \nabla_{\textbf{r}} \psi^*(\textbf{r},t) -\psi^*(\textbf{r},t)\nabla_{\textbf{r}}\psi(\textbf{r},t)],
\end{equation}
\begin{equation} \label{eq26}
 \textbf{B}(\textbf{r},t)=\frac{\mu_0}{4\pi}\int [\frac{\textbf{j}(\textbf{r}',t_r)}{|\textbf{r}-\textbf{r}'|^3}+\frac{1}{|\textbf{r}-\textbf{r}'|^2c}\frac{\partial \textbf{j}(\textbf{r}',t')}{\partial t}]\times (\textbf{r}-\textbf{r}')d^3\textbf{r}'.
\end{equation}
$t_r$ is the retarded time $t-r/c$ and $\mu_0=4\pi \times 10^{-7}$ NA$^{-2}$ (6.692$\times10^{-4}$ a.u.). For the static time-independent conditions occurring after the laser pulse, then Eq. (\ref{eq25}) reduces to $\displaystyle \textbf{B}(\textbf{r})=\frac{\mu_0}{4\pi}\int \frac{\textbf{j}(\textbf{r}')\times (\textbf{r}-\textbf{r}')}{|\textbf{r}-\textbf{r}'|^3}d^3\textbf{r}'$ in accord with the classical Biot-Savart law\cite{zzz}.

\begin{table}[!t ] \centering
\tbl{Maximum local $B_{max}(\textbf{r},t)$ and total volume magnetic field ${B}(t)$ and currents $j(t)$ by circularly polarized attosecond UV pulses with intensity $I=2\times10^{16}$ W/cm$^2$, wavelength $\lambda=50$ nm, and duration $3\tau=500$ as at different times.}
{\begin{tabular} {cccccccc}
\hline
& units &1.0$\tau$ & 1.25$\tau$ &1.5$\tau$ &1.625$\tau$& 1.75$\tau$& 2.0$\tau$\\ \hline

 {$B_{max}(\textbf{r},t)$} & T & 6.242 &  12.285 & 11.971 & 9.121 & 6.505& 4.167  \\

 {$B(t)$} & T$\cdot a_0^3$ & 6.670 &  17.612 & 27.797&29.300 & 27.517& 17.320 \\

 {$j(t)$}  & fs$^{-1}$$\cdot a_0$ & 5.209 & 12.568 & 10.625 & 10.211 &10.294 &8.351\\
  & mA$\cdot a_0$ & 0.834 &  2.011 &  1.700 & 1.634 &1.674 & 1.336 \\
 \hline
\end{tabular}}\label{tab1}
\end{table}

We calculate the maximum local $B_{max}(\textbf{r},t)$ and total volume average attosecond magnetic fields and the corresponding currents by integrating $B(t)=|\int\textbf{B}(\textbf{r},t)d\textbf{r}^3|$ and $j(t)=|\int\textbf{j}(\textbf{r},t)d\textbf{r}^3|$ over the electron $\textbf{r}$ space. Table \ref{tab1} lists values of $B_{max}(\textbf{r},t)$, $B(t)$, and  $j(t)$ at different moments, illustrated in Fig. \ref{fig2}(b). Both $B(t)$ (in units of T$\cdot a_0^3$, where $a_0$ is Bohr radius) and $j(t)$ (fs$^{-1}\cdot a_0$) vary with time, increasing first and then decreasing in phase with the pulse. From Tab. \ref{tab1} one obtains that the maximum total volume magnetic field is induced at $t=1.625\tau$=270 asec with strength $B=1.172\times10^{-4}$ a.u. =29.3 T$\cdot a_0^3$ (2.93$\times10^5$ Gauss$\cdot a_0^3$). The maximum local magnetic field $B_{max}(\textbf{r},t)=12.285$ T and the maximum electronic current $j=0.304$ a.u.=12.568 fs$^{-1}$$\cdot a_0$ (2.011 mA$\cdot a_0$) is produced at time $t=1.25\tau=207$ asec, where $E_x=0$ and $E_y=-E$ (Fig. 1).
A time delay $\Delta t=0.375\tau=63$ asec occurs between maximum current $j(t=1.25\tau)=12.568$ fs$^{-1}$$\cdot a_0$ and $B(t=1.625\tau)=29.3$ T$\cdot a_0^3$ (Table \ref{tab1}).

 The current electron trajectories are functions of the pulse wavelength $\lambda$ (frequency $\omega$) and duration $n_l\tau$ \cite{32}. We explain this from the classical model \cite{32}, a generalization of the linear polarization model \cite{14}. Assuming the zero initial electron velocities $\dot x(t_0)=\dot y(t_0)=0$, where $t_0$ is the ionization time, the induced time dependent velocities are \begin{eqnarray} \label{VC} \displaystyle { \begin{array}{l}
\dot x(t)=\displaystyle -{E}/{\omega} \left( \sin\omega
t-\sin\omega t_0 \right ),  \\ \dot y(t)=\displaystyle
-{E}/{\omega} \left( \cos\omega t_0 -\cos \omega t
 \right ),\end{array} }
\end{eqnarray}
The corresponding displacements
are
\begin{eqnarray} \label{DC} \displaystyle { \begin{array}{l}
 x(t)=\displaystyle -{E}/{\omega^2} \left[\cos\omega t_0-\cos\omega t-(\omega t-\omega t_0)\sin\omega t_0 \right ],
\\y(t)=\displaystyle -{E}/{\omega^2} \left[\sin\omega
t_0-\sin\omega t+ (\omega t-\omega t_0)\cos\omega t_0
 \right ],\end{array} }
\end{eqnarray}
with $x(t_0)=y(t_0)=0$ corresponding to recollision at the center of the molecule ($r=0$). From Eqs. (\ref{VC}) and (\ref{DC}) it is found that increasing the pulse wavelength $\lambda$ leads to increase of the maximum induced electron velocity $v=2E/\omega$ at $\omega
t-\omega t_0=(2n'+1)\pi$, and the corresponding radii \begin{equation} r_{n'}={2E}/{\omega^2} \left [
1+(n'+{1}/{2})^2\pi^2 \right ]^{1/2},\end{equation} $n'=0,1,2,\cdots$. For a moving point charge the corresponding classical magnetic field can be expressed as \begin{equation} \label{BC}
\textbf{B}=\frac{\mu_0}{4\pi}\frac{\textbf{v}\times\textbf{r}}{r^3}.\end{equation} From the above equations one then gets the maximum field $B\sim v/r^2_{n'}\sim \omega^3/E$ at times $t_0+(2n'+1)\pi/\omega$.
Therefore, an increase of $\lambda$ or lower $\omega$, results in a decrease of the magnetic field due to large radii $r_{n'}$ of the electron, thus reducing the efficiency of the attosecond magnetic field generation. Longer pulse duration have a similar effects. Of note is that we only present the results for the single photon ionization processes. Increasing the pulse frequency results in decrease of the ionization rate in spite of decrease of the radii of the free photoelectron. Consequently, weaker magnetic field pulses are nevertheless produced.

\section{Summary}
Molecular high-order harmonic generation (MHOHG) is generally an efficient source of coherent radiation by molecules subjected to intense ultrashort ionizing laser pulses. The nonspherical symmetry of molecules eliminates selection rules in spherical atoms. Nevertheless, in both atomic and molecular spectra, recollision of an ionized electron with its parent ion is the essential process for harmonic generation. Circularly polarized intense laser pulses inhibit in general recollision with parent ions, so that combinations of different polarization pulses suggest obvious general schemes for recollision in order to produce circularly polarized harmonics. High order generation of circularly polarized harmonics is thus an essential route for producing circularly polarized attosecond pulses. Schemes involving \textit{co-rotating} and \textit{counter-rotating} intense circularly polarized pulses are analyzed and shown to lead to recollision in general. Co-rotating pulses result in large Coriolis forces as opposed to counter-rotating combinations of pulses which favour high frequency recollisions. In the limit of near equal frequency co-rotating and counter-rotating pulses, one obtains a scheme involving high frequency pulses in combinations with low frequency TeraHertz pulses. Such a combination has been shown to produce copious circularly polarized harmonics for the generation of attosecond pulses \cite{38,39}. The theoretical analysis presented in this paper shows that in principle, one can also use static magnetic fields for controlling Coriolis forces which reduce the efficiency of circularly polairzed MHOHG.Circulalry polarized attosecond pulses provide new tools for creating in matter attosecond coherent electronic currents, which are sources of attosecond magnetic field pulses. These new tools should lead to important advances in the study of magnetic light-matter interaction, in particular in quantifying the magnetic nature of light absorption-emission in ultrafast femto-attosecond nano-optics and spintronics \cite{x,xx}.

\section{Acknowledgments}
The authors thank RQCHP and Compute Canada for access to massively parallel
computer clusters and NSERC, FQRNT for financial support of this research in the
ultrafast science programs.

\end{document}